\documentclass[%
preprint%
,secnumarabic%
,tightenlines%
,amssymb, amsmath,nobibnotes, aps,nofootinbib,showpacs,]{revtex4}
\usepackage{epsfig}%
\usepackage{graphicx}%
\expandafter\ifx\csname package@font\endcsname\relax\else
 \expandafter\expandafter
 \expandafter\usepackage
 \expandafter\expandafter
 \expandafter{\csname package@font\endcsname}%
\fi

\begin{document}

\title{Using Weyl Symmetry to make Graphene  \\ a real lab for Fundamental Physics\footnote{Invited talk at the ``3rd O'Raifeartaigh Conference on Symmetry and Integrability'' 19-21 July, 2012 - Munich, Germany.}}
\author{Alfredo Iorio}
\affiliation{Institute of Particle and Nuclear Physics \\ Faculty of Mathematics and Physics \\ Charles University \\ Prague, Czech Republic\\ \\
alfredo.iorio@mff.cuni.cz}

\def\be{\begin{equation}}
\def\ee{\end{equation}}
\def\al{\alpha}
\def\bea{\begin{eqnarray}}
\def\eea{\end{eqnarray}}

\begin{abstract}
In the first attempt to introduce gauge theories in physics, Hermann Weyl, around the 1920s, proposed certain scale transformations to be a fundamental symmetry of nature. Despite the intense use of Weyl symmetry that has been made over the decades, in various theoretical settings, this idea never found its way to the laboratory. Recently, building-up from work by Lochlainn O'Raifeartaigh and collaborators on the Weyl-gauge symmetry, applications of Weyl-symmetry to the electronic properties of graphene have been put forward, first, in a theoretical setting, and later, in an experimental proposal. Here I review those results, by enlarging and deepening the discussion of certain aspects, and by pointing to the steps necessary to make graphene a testing ground of fundamental ideas.
\end{abstract}

\pacs{11.30.-j, 04.62.+v,  72.80.Vp}


\maketitle

\section{Recollection of Lochlainn O'Raifeartaigh and Weyl symmetry}

\begin{figure}
\centering \leavevmode \epsfxsize=4cm \epsfysize=5cm
\epsffile{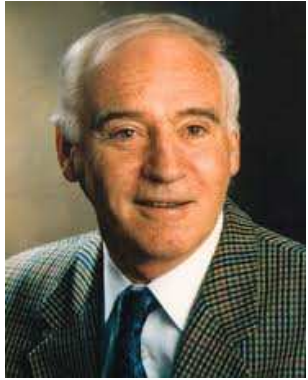}
\caption{The ``official'' picture of Lochlainn O'Raifeartaigh.} \label{picture1}
\end{figure}

Since I had the privilege to be the last PhD student of Lochlainn O'Raifeartaigh (LOR), and since the way I achieved it is related to the matter discussed here, it is perhaps worth to recollect that story in a conference in memory of LOR. Indeed, it is yet another anecdote about the rare human qualities of this first rank scientist, and, I believe, this is also an important aspect of his legacy.

In the last years of his ground-breaking career, LOR (see Fig.1) was way too advanced to have the time to rise-up PhD students. The youngest collaborators he had been admitting in DIAS were smart post-docs. Thus, when a stubborn, no-one-word-of-proper English character came along in his office in 1996, to propose himself for a PhD with ``\textit{Professore Oraifferti}'' as supervisor, LOR was probably disoriented. But LOR was not a person to give-up for matters related to cultural barriers, so he handed over to me Bailin and Love's book on gauge theories, and said, slowly articulating the words: ``\textit{This is what I do}''.

A few weeks later, as I came back to him, LOR gave me a more important chance. He had me seat in front of his office's desk in DIAS (see Fig.2, not a true picture of his hands, but a faithful reproduction of what I used to see for a few months). He showed to me a draft-paper by himself, Ivo Sachs and Chris Wiesendanger, on when scale invariance implies full conformal invariance. There local Weyl symmetry plays a crucial role. The paper was nearly complete, and he said (again, articulating the words one by one): ``\textit{Apply this to fields of any spin, and spacetime dimensions}''. LOR meant that to be an MSc thesis work, but the assignment was finished in a couple of weeks only, for the surprise of LOR (and for the extra work of the two brilliant post-doc coauthors, who had to include my hand-written results in a revised version of the paper, and one more author, that is Ref.\cite{lor}). With this, LOR eventually gave me the most valuable opportunity he could give me, and accepted me as his PhD student.

\begin{figure}
\centering \leavevmode \epsfxsize=5cm \epsfysize=4cm
\epsffile{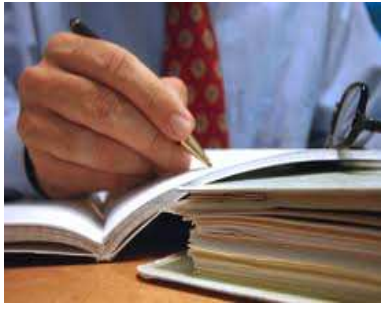}
\caption{My view of Professor O'Raifeartaigh as a perspective PhD student (1996).} \label{picture2}
\end{figure}

\begin{figure}[h]
\begin{minipage}{14pc}
\includegraphics[height=.2\textheight, width=12pc]{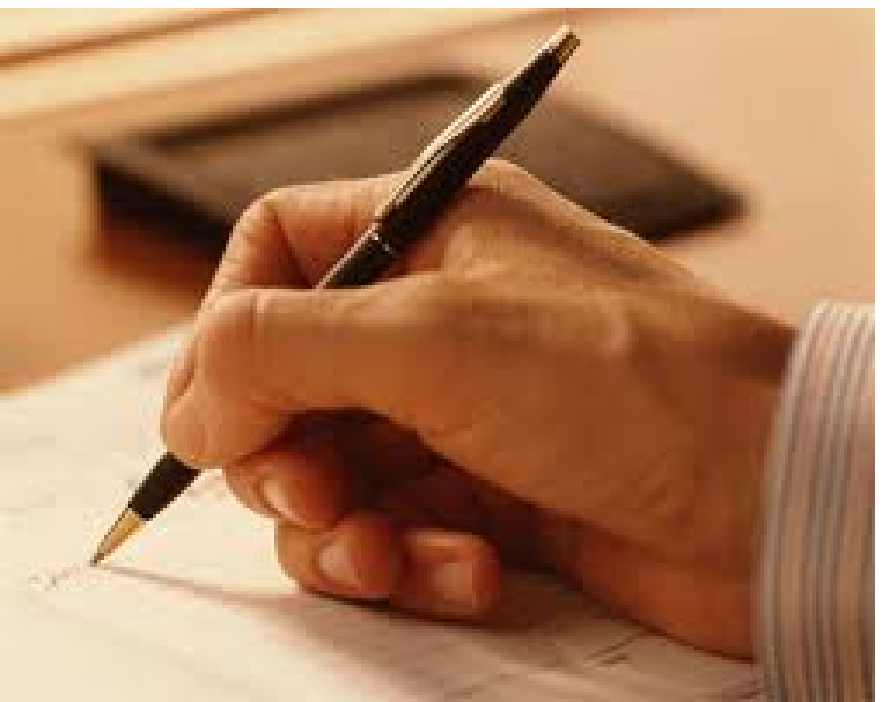}
\end{minipage}
\begin{minipage}{14pc}
\includegraphics[height=.2\textheight, width=12pc]{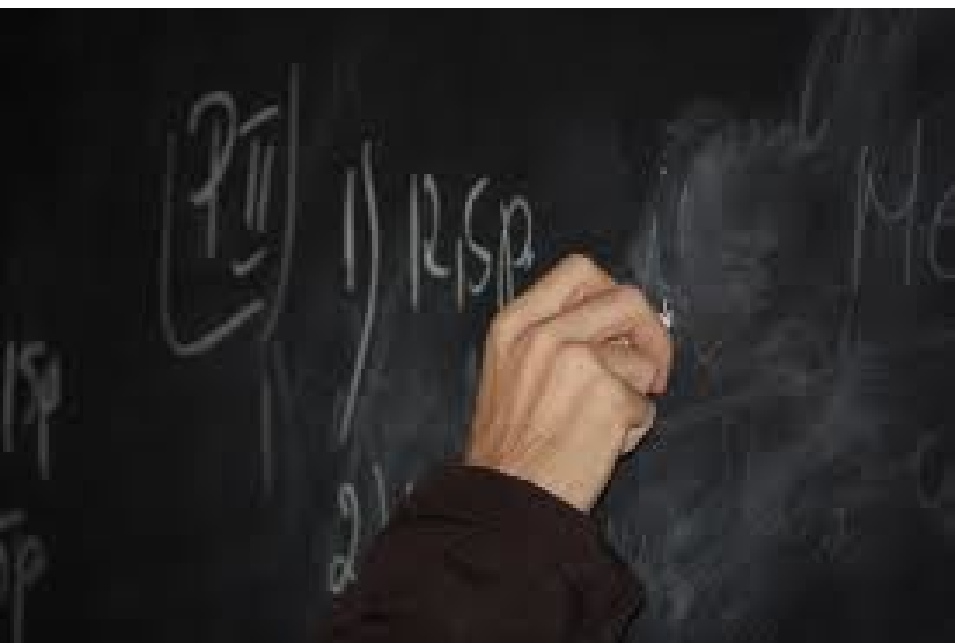}
\end{minipage}
\caption{Improvement of the angle of view, gained over the PhD years (1996--98), and ``promotion'' to the black-board (1999--2000).}
\end{figure}

The recollection could go on for a long while, but then it would depart too much from a scientific paper, so I leave it to Fig.~3 (as above, not original pictures, but faithful reproductions of the images I remember). Let me then move on to physics, by first summarizing here what, in the occasion described above, I learned to be Weyl symmetry.

Suppose that the following action for fields of any spin, at most quadratic in the derivatives of the fields
\begin{eqnarray}
    A (\Phi_i) = \int d^n x {\cal L} (\Phi_i , \partial_a \Phi_i) \;,
\end{eqnarray}
(here $A$ refers to Minkoswki spacetime, while $\cal A$ refers to a generic $g_{\mu \nu}$, not necessarily curved) is symmetric under rigid ($\sigma_a \equiv \partial_a \sigma = 0$) scaling
\begin{eqnarray}
    x^a \to e^\sigma x^a \;\;  {\rm and} \;\; \Phi_i \to e^{d_\Phi \sigma} \Phi_i  \;.
\end{eqnarray}
That is a spatiotemporal symmetry, but, due to invariance under diffeomorphisms, ${\cal A} (\Phi_i)$ is symmetric under\footnote{Here our notations:
\[
{\cal A} (\Phi_i) \equiv \int d^n x \sqrt{g} {\cal L} (\Phi_i , \nabla_a \Phi_i)
\]
$\mu , \nu = 0, 1, ..., n-1$ are Einstein indices, $a, b = 0, 1, ..., n-1$ are Lorentz indices, $i,j$ are spin indices,
$\nabla_\mu \Phi_i = \partial_\mu \Phi_i + {\Omega_\mu}_i^{\; j} \Phi_j$ with $\nabla_a = E_a^\mu \nabla_\mu$, ${\Omega_\mu}_i^{\; j} =  \frac{1}{2} \omega_\mu^{a b} (J_{a b})_i^{\; j}$, $(J_{a b})_i^{\; j}$ are the Lorentz generators, and
${\omega_\mu}^a_{\; b} = e^a_\lambda (\delta^\lambda_\nu \partial_\mu + \Gamma_{\mu \nu}^\lambda) E^\nu_b$ is the
spin connection coming from the metricity condition
$\nabla_\mu e^a_\nu = \partial_\mu e^a_\nu - \Gamma_{\mu \nu}^\lambda e^a_\lambda +  \omega_{\mu \; b}^a e^b_\nu = 0$. We also introduced the Vielbein $e^a_\mu$ and its inverse $E_a^\mu$, satisfying $\eta_{a b} e^a_\mu e^b_\nu = g_{\mu \nu}$, $e^a_\mu E_a^\nu = \delta_\mu^\nu$, $e^a_\mu E_b^\mu = \delta_b^a$, where $\eta_{a b} = {\rm diag} (1, -1, ...)$. For more notations see \cite{iorio}.}
\begin{eqnarray}
     x^\mu \to x^\mu \; , \; e^a_\mu \to e^\sigma e^a_\mu \;\; {\rm and} \;\; \Phi_i \to e^{d_\Phi \sigma} \Phi_i \label{localWeyl} \;,
\end{eqnarray}
hence rigid Weyl transformations can be seen as {\it internal} transformations (i.e., they involve only the fields, the metric and $\Phi_i$, not the coordinates).

As any other internal symmetry, also Weyl's can be gauged. First one promotes (\ref{localWeyl}) to a local transformation, i.e. $\sigma \Rightarrow \sigma(x)$, then one promotes the original action to a ``Weyl-gauged'' one, 
${\cal A} (\Phi_i , \nabla_a \Phi_i) \Rightarrow {\cal A}_W (\Phi_i , {\cal D}_a \Phi_i)$, 
where
${\cal D}_\mu \Phi_i = \nabla_\mu \Phi_i + (\Lambda_\mu^{\; \nu})_i^{\; j} W_\nu \Phi_j$, 
with Weyl field and derivative responding to (\ref{localWeyl}) as
\begin{eqnarray}
    W_\mu \to W_\mu - \sigma_\mu \; \; {\rm and} \; \; {\cal D}_\mu \Phi_i \to e^{d_\Phi \sigma} {\cal D}_\mu \Phi_i \label{WfieldDtrf}\;,
\end{eqnarray}
respectively. Here I used here the virial tensor, introduced in \cite{Callan:1970ze}
\begin{eqnarray}
    (\Lambda_{\mu \nu})_i^{\; j} = d_\Phi g_{\mu \nu} \delta_i^{\; j} + (J_{\mu \nu})_i^{\; j} \;.
\end{eqnarray}
At the end of this (standard) procedure, the transformations (\ref{localWeyl}), alongside with (\ref{WfieldDtrf}), are indeed a symmetry of the new action: ${\cal A}_W (\Phi_i) \to {\cal A}_W (\Phi_i)$.

The status of the massless Dirac action in any dimension is very special because\footnote{For the moment, we shall take $\hbar = 1 = v$ with $v$ a parameter with the dimensions of velocity, for graphene it is $v = v_F$, the Fermi velocity, for truly relativistic Dirac it is, of course, $v = c$.}
\begin{equation}\label{diracgeneral}
     {\cal A}_W (\Psi) = i \int d^n x \sqrt{g} \; \bar{\Psi} \gamma^a E^\mu_a
     (\nabla_\mu + \Lambda_\mu^{\; \nu} W_\nu) \Psi = {\cal A} (\Psi) \;,
\end{equation}
due to $\gamma^\mu \Lambda_{\mu \nu} = d_\Psi \gamma_\nu + \gamma^\mu J_{\mu \nu} = 0$, where I used $d_\Psi = (1 -n)/2$ and the definition of the Lorentz generators $J_{\mu \nu} = \frac{1}{4} [\gamma_\mu, \gamma_\nu]$. This means that not only the Dirac action in flat space
\begin{equation}\label{diracflat}
    A (\Psi) = i \int d^n x \bar{\Psi} \gamma^a \partial_a \Psi \;,
\end{equation}
is invariant under the full conformal group, $SO(n,2)$, but that the curvilinear, or truly curved spacetime, action
\begin{equation}\label{diraccurve}
 {\cal A} (\Psi) = i \int d^n x \sqrt{g} \; \bar{\Psi} \gamma^a E^\mu_a (\partial_\mu + \frac{1}{2} \omega_\mu^{\; b c} J_{b c}) \Psi \;,
\end{equation}
is locally Weyl invariant. This can be also seen directly by transforming (\ref{diraccurve}) under (\ref{localWeyl}) with $\sigma(x)$, obtaining
\begin{equation}\label{directweyl}
    {\cal A} (\Psi) \to {\cal A} (\Psi) + i \int d^n x \sqrt{g} \; \bar{\Psi} \gamma^\mu \Lambda_\mu^{\; \nu} \sigma_\nu \Psi = {\cal A} (\Psi) \;.
\end{equation}
Here it was used $\omega_\mu^{\; a b} \to \omega_\mu^{\; a b} + (e^a_\mu e^b_\nu - e^b_\mu e^a_\nu)$.

\section{Weyl symmetry of the Dirac action and graphene}

In any dimension Weyl's is a very large symmetry that ${\cal A} (\Psi)$ enjoys because there are no restrictions to $\sigma(x)$ of any sort. This happens for the action (\ref{diraccurve}) as it stands, which means that the theory is, we may say, ``intrinsically gauged'': ${\cal A}_W (\Psi) = {\cal A}(\Psi)$. In what follows I shall try to exploit this symmetry for graphene. There a crucial role is played by the dimensionality of the problem, that is $n=3$, but let us see here what this symmetry means in physical terms in any dimensions \cite{iorio}.

When two metrics are related as $g'_{\mu \nu} = e^{2 \sigma (x)} g_{\mu \nu}$ the classical physics for the field $\Psi' = e^{d_\Psi \sigma (x)} \Psi$ in $g'_{\mu \nu}$ is precisely the same as the classical physics for the field $\Psi$ in $g_{\mu \nu}$  because ${\cal A}(g_{\mu \nu}, \Psi , \nabla_a \Psi) = {\cal A}({g'}_{\mu \nu}, \Psi' , {\nabla'}_a \Psi')$. Furthermore, we can choose
\begin{equation}
g'_{\mu \nu} \equiv \eta_{\mu \nu} \;,
\end{equation}
and obtain
\begin{equation}\label{corerelation1}
i \int d^n x \sqrt{g} \bar{\Psi} \gamma^a E_a^\mu \nabla_\mu \Psi = i \int d^n x \; e^{- (n -1) \sigma} \; \bar{\Psi} \gamma^a
(\partial_a - \frac{n - 1}{2} \sigma_a) \Psi = i \int d^n x \bar{\Psi}' \gamma^a \partial_a \Psi' \;,
\end{equation}
when
\begin{equation}\label{corerelation2}
    g_{\mu \nu} = e^{-2 \sigma (x)} \eta_{\mu \nu} \;\;\;\; {\rm and} \;\;\;\; \Psi = e^{\frac{(n-1)}{2} \sigma(x)} \Psi' \;,
\end{equation}
where we used the simple algebra illustrated in \cite{iorio}.

In (\ref{corerelation1}) we are dealing with a conformally invariant field in a conformally flat spacetime (a case sometimes referred to as {\it conformal triviality} \cite{birrellanddavies}, a name that emphasizes the simplest possible case of QFT in curved space, but, perhaps, does not justice to the fact that all the key features are indeed at work, see later). This means that, if the spacetime is only curved in a conformally flat fashion, the effects of curvature are null on the classical physics of a massless Dirac field: the classical physics is, once and for all, governed by the flat space action.

Since we have in mind to see Weyl symmetry emerging in the case of graphene, the above is of particular importance for us. In a moment, I shall remind why the flat space $n=3$ Dirac massless action is experimentally sound in the description of certain regimes of graphene's conductivity properties, for the moment refer to \cite{pacoreview2009}. If one then accepts this flat space description, one should also consider that the same action describes (in principle) infinitely many other curved configurations. Due to the current interest on how curvatures/deformations change the properties of graphene, this is also of importance on the condensed matter side.

How about the quantum case? We are free to choose the approach we like to take into account how the quantum effects have to be included. One obvious thing to focus on would be the trace anomalies, that here could be due to the introduction of curvature, hence of a length scale. Nonetheless, for the use I have in mind (I shall be dealing with an odd-dimensional spacetime) trace anomalies will not play a role\footnote{Perhaps, they might become important at a later stage, i.e. when a dimensional reduction would bring us into an even-dimensional spacetime. We are currently exploring this direction in relation to the opening of a gap.}, (see, e.g., \cite{chr} for explicit computations of the absence of anomalies in odd dimensions).

One way to bring to the surface the quantum effects, is to take into account how the quantum vacua encode the geometry of the different spacetimes through a condensate structure. This is a customary approach in QFT in curved spacetimes, see, e.g., \cite{TFDBH, birrellanddavies}. Consider, e.g., the positive frequency Wightman  two-point function for the flat (primed) fields and vacuum \cite{birrellanddavies}
\begin{eqnarray}
    S ' (x_1, x_2) \equiv \; '\langle 0| \Psi ' (x_1) \bar{\Psi}' (x_2) |0 \rangle ' \;.
\end{eqnarray}
The operator implementing Weyl transformations quantum mechanically $\Psi (x) = U \Psi' (x) U^{-1} = e^{\frac{n-1}{2} \sigma(x)} \Psi'(x)$ is
\begin{eqnarray}
    U = \exp \left\{ \frac{1 - n}{2} \int d^n y \sigma(y) {\Psi'}^\dagger (y) \Psi'(y) \right\} \label{U} \;,
\end{eqnarray}
hence the transformed vacuum, $|0\rangle = U |0\rangle'$, reveals the typical condensate structure in a curved background\footnote{In fact, this is a feature common to a variety of other cases, both in high energy and in condensed matter. It takes place when there is more than one Hilbert/Fock space available to describe the system, from physically different ``frames of reference'' (an expression used here in the broad sense). On the general issue see, e.g., \cite{vitQFTbook, habilitation}, and, for an application to supersymmetry breaking, see \cite{susymix}.}.

We can consider either (unprimed objects refer to the curved case)
\begin{eqnarray}
    S  (x_1, x_2) & \equiv & \langle 0| \Psi  (x_1) \bar{\Psi} (x_2) |0 \rangle
    = \; '\langle 0| U^{-1} U \Psi ' (x_1) U^{-1} U \bar{\Psi}' (x_2) U^{-1} U |0 \rangle ' \nonumber \\
    &=&  \; '\langle 0| \Psi ' (x_1) \bar{\Psi}' (x_2) |0 \rangle '  = S ' (x_1, x_2) \label{directgreen2} \;,
\end{eqnarray}
or
\begin{eqnarray}
    S^\sigma (x_1, x_2) \equiv \; '\langle 0| \Psi  (x_1) \bar{\Psi} (x_2) |0 \rangle ' & = & e^{\frac{n-1}{2} (\sigma(x_1) + \sigma(x_2))} \;
    S ' (x_1, x_2)  \label{reversegreen1} \\
    S^{- \sigma} (x_1, x_2) \equiv \; \langle 0| \Psi ' (x_1) \bar{\Psi}' (x_2) |0 \rangle & = & e^{\frac{1 - n}{2} (\sigma(x_1) + \sigma(x_2))} \;
    S ' (x_1, x_2) \label{reversegreen2} \;,
\end{eqnarray}
depending on how we perform the measurements. In the first case, Eq. (\ref{directgreen2}), the measurements are made within the frame of reference that is co-moving with the particles described by the field $\Psi$, hence the effects of curvature are completely removed also at the quantum level. In the other cases, the measurements are made by an observer that sees the quantum effects of curvature in the form of a condensate in the vacuum\footnote{The quantum effects of the classical symmetry can also be appreciated within the standard path integral formalism
\begin{eqnarray}
S^{PI} (x_1, x_2) & \equiv & \frac{\int {\cal D} \Psi {\cal D} \bar\Psi \; \Psi(x_1) \bar\Psi(x_2) \; \exp\{ i {\cal A}(\Psi) \}}
{\int {\cal D} \Psi {\cal D} \bar\Psi \exp\{ i {\cal A}(\Psi) \}}  \label{PIGreen1} \\
& = & e^{\frac{n-1}{2} (\sigma(x_1) + \sigma(x_2))} \frac{\int {\cal D} \Psi' {\cal D} \bar\Psi' \; \Psi'(x_1) \bar\Psi'(x_2) \; \exp\{ i {\cal A}(\Psi') \}}{ \int {\cal D} \Psi' {\cal D} \bar\Psi' \exp\{ i {\cal A}(\Psi') \}}  \label{PIGreen2} \\
& \equiv & e^{\frac{n-1}{2} (\sigma(x_1) + \sigma(x_2))} S' (x_1, x_2) \label{PIGreen3} \;.
\end{eqnarray}
Here we exploited the classical Weyl-symmetry of the action ${\cal A}(\Psi) = {\cal A}(\Psi')$, and, since $\sigma(x)$ is an external field, we used ${\cal D} \Psi = e^{\frac{n-1}{2} \sigma(x)} {\cal D} \Psi'$, similarly for ${\cal D} \bar\Psi$, thus, in the ratio, the overall factor $e^{\frac{n-1}{2} (\sigma(x) + \sigma(y))}$ cancels. From the above we see that $S^{PI} \equiv S^\sigma$. This is so because we can as well read (\ref{PIGreen2}) as
\begin{equation}\label{linkgreen}
    \frac{\int {\cal D} \Psi' {\cal D} \bar\Psi' \; \Psi(x_1) \bar\Psi(x_2) \; \exp\{ i {\cal A}(\Psi') \}}{\int {\cal D} \Psi' {\cal D} \bar\Psi' \exp\{ i {\cal A}(\Psi') \}} \equiv  \; '\langle 0| \Psi  (x_1) \bar{\Psi} (x_2) |0 \rangle ' \;.
\end{equation}
After having exploited the classical symmetries in the path integral Green's functions, we can then use the operator approach to find the $U$s corresponding to these symmetries, i.e. for us here the $U$s in (\ref{U}), and we can as well construct the invariant Green's function as in (\ref{directgreen2}).}.

\begin{figure}
 \centering
  \includegraphics[height=.2\textheight]{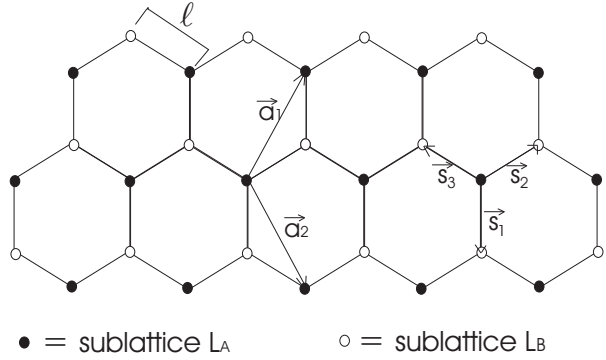}
  \caption{The honeycomb graphene lattice.}
\label{honeycomb}
\end{figure}

It is time to see why I claim that the above structures naturally emerge in the physics of graphene \cite{iorio}. Graphene is a two-dimensional honeycomb lattice of carbon atoms arranged in two triangular sub-lattices, say $L_A$ and $L_B$, whose electrons in the $\pi$-bonds belonging to one sublattice can hop to the nearest neighbor sites of the other sublattice, see Fig.~\ref{honeycomb}. The electronic properties of graphene are ascribed to these electrons. The elastic properties, instead, are ascribed to the $\sigma$-bonds, and involve an energy orders of magnitude stronger than that relative to the $\pi$-bonds. The material was produced in a laboratory only in 2004 \cite{geimnovoselovFIRST}, but was theoretically studied, as a sort of toy model, in the 1940s \cite{wallace}, and in the 1980s, more explicitly as for the Dirac-like properties, in \cite{semenoff}. Let me briefly show how such Dirac-like structure naturally emerges.

The electronic properties of this material, in the tight-binding low-energy approximation, are customarily described by the Hamiltonian
\begin{eqnarray}
    H = - t \sum_{\vec{r} \in L_A} \sum_{i =1}^3 \left( a^\dagger (\vec{r}) b(\vec{r} +\vec{s}_i)
    + b^\dagger (\vec{r} +\vec{s}_i) a (\vec{r}) \right) \;,
\end{eqnarray}
where $\ell \simeq 2.5${\AA}, $t \simeq 2.7$~eV, $a, a^\dagger, b, b^\dagger$ are the anti-commuting operators for the planar ($\vec{r} = (x,y)$, $\vec{s}_i = (s_i^x, s_i^y)$) electrons, and, see Fig.~\ref{honeycomb}, $\vec{a}_1 = \frac{\ell}{2} (\sqrt{3}, 3)$, $\vec{a}_2 = \frac{\ell}{2} (\sqrt{3}, - 3)$, $\vec{s}_1 = \ell (0, - 1)$, $\vec{s}_2 = \frac{\ell}{2} (\sqrt{3}, 1)$, $\vec{s}_3 = \frac{\ell}{2} (-\sqrt{3}, 1)$.

If we Fourier transform, $a(\vec{r}) = \sum_{\vec{k}} a(\vec{k}) e^{i \vec{k} \cdot \vec{r}}$, etc,  then $H = \sum_{\vec{k}} ( f(\vec{k}) a^\dagger (\vec{k}) b(\vec{k}) + {\rm h.c.} )$, (here $\ell = 1$), with
\begin{equation}
  f(\vec{k}) = - t e^{-i k_y} \left( 1 + 2 e^{i \frac{3}{2} k_x} \cos(\frac{\sqrt{3}}{2} k_x) \right) \;.
\end{equation}
Solving $E(\vec{k}) = \pm |f (\vec{k})| \equiv 0$, gives the modes with zero one-particle energy , i.e. it tells us where, in the Brillouin zone, conductivity and valence bands touch (if they do). Indeed, for graphene, this happens, pointing to a massless (read: gapless) spectrum. Furthermore, the solution is not a Fermi \textit{line} (the $n=3$ version of the Fermi surface encountered in $n=4$), but rather they are two Fermi \textit{points}, $\vec{k}^D_\pm = (\pm \frac{4 \pi}{3 \sqrt{3}}, 0)$. The superscript ``$D$'' stands for ``Dirac'', and that refers to the last and more important surprise here: near those points (they are actually six, but only the indicated two are inequivalent) the spectrum is linear. It is like for a relativistic theory, whereas, in such a non-relativistic setting, one would expect quadratic dispersion relations.

The pseudo-relativistic structure is seen even more cleanly when one linearizes around $\vec{k}^D_\pm$: $\vec{k}_\pm \simeq \vec{k}^D_\pm + \vec{p}$, and $f_\pm (\vec{p}) \equiv f (\vec{k}_\pm) = \pm \frac{3 t}{2} (p_x \pm i p_y)$, $a_\pm (\vec{p}) \equiv a (\vec{k}_\pm)$,  $b_\pm (\vec{p}) \equiv b (\vec{k}_\pm)$, then
\begin{eqnarray}
    H|_{\vec{k}_\pm} & \simeq & \sum_{\vec{p}} [ f_+ a_+^\dagger b_+ +
f_- a_-^\dagger b_- + f^*_+ b_+^\dagger a_+ + f^*_- b_-^\dagger a_- ] (\vec{p}) \\
    & = & v_F  \sum_{\vec{p}} [ (a_+^\dagger , b^\dagger_+) \left(\begin{array}{cc} 0 & p_x + i p_y \\ p_x - i p_y & 0 \\ \end{array} \right)
\left(\begin{array}{c} a_+ \\ b_+ \end{array}\right)  \\
& - & (a_-^\dagger , b^\dagger_-) \left(\begin{array}{cc} 0 & p_x + i p_y \\ p_x - i p_y & 0 \\ \end{array} \right)
\left(\begin{array}{c} a_- \\ b_- \end{array}\right) ] (\vec{p})
\end{eqnarray}
where $v_F = 3 t / 2$ is the Fermi velocity. Thus we see spin matrices and the Dirac spinors emerging entirely from the lattice structure. More explicitly, the Hamiltonian can be written as
\begin{eqnarray}
    H  &=&  v_F \sum_{\vec{p}} \left(\psi_+^\dagger \vec{\sigma} \cdot \vec{p} \; \psi_+
    + \psi_-^\dagger \vec{\sigma}^* \cdot \vec{p} \; \psi_- \right) \label{Hwithk} \\
    & = & - i v_F \int d^2 x \left( \psi_+^\dagger \vec{\sigma} \cdot \vec{\partial} \; \psi_+
    + \psi_-^\dagger \vec{\sigma}^* \cdot \vec{\partial} \; \psi_- \right) \label{HGrapheneB}
\end{eqnarray}
with $\vec{\sigma} \equiv (\sigma_1, \sigma_2)$, $\vec{\sigma}^* \equiv (-\sigma_1, \sigma_2)$, $\sigma_i$ the Pauli matrices, and $\psi_\pm \equiv \left( a_\pm , b_\pm \right)^{\rm T}$ are Dirac spinors. In the step from (\ref{Hwithk}) to (\ref{HGrapheneB}) we not only moved back to configuration space, but we also used the continuum approximation. For energies small compared to $\hbar v_F / \ell \sim 2.6$~eV (that numerically coincides with the value of the hopping parameter $t \sim 2.7$~eV), this approximation will well describe the physics, and this is precisely the range of energies where the whole of the illustrated Dirac construction works.

Now, since we do not consider phenomena mixing the two Fermi points, let us focus on a {\it single} one, say, $\psi \equiv \psi_+$. If we want to take seriously these pseudo-relativistic features, one important conceptual step to take is to consider an action, rather than a Hamiltonian \cite{iorio}
\begin{equation}
    A =  \int d^3 x (i \psi^\dagger \; \dot{\psi} - {\cal H}) = i \int d^3 x \bar{\psi} \; \gamma^a \partial_a \; \psi
\end{equation}
where $v_F = 1$, and the $\gamma$ matrices, $\gamma^0 = \sigma_3$, $\gamma^1 = i \sigma_2$, $\gamma^2 = - i \sigma_1$, satisfy all the standard properties.

To use the power of Weyl symmetry, we need to be even more serious with time, and, when the graphene sheet is curved, assume that the metric experienced by these pseudoparticles is
\begin{equation}
g^{(3)}_{\mu \nu} (x,y) = \left(\begin{array}{cc} 1 &  \\  & g^{(2)}_{\alpha \beta} (x,y)  \end{array} \right) \;,
\end{equation}
where the spatial curvature is all in $g^{(2)}_{\alpha \beta}$.

Here I need to make an important comment. When the deformation is inelastic, say by making defects, the metric is taking into account the intrinsic curvature. On the other hand, the metric is also able to include extrinsic curvature effects, by, e.g., having zero Riemannian (intrinsic) curvature, but nonzero connections. With this in mind, the action to consider is then
\begin{equation}
{\cal A} = i \int d^3 x \sqrt{g^{(3)}} \; \bar{\psi} \gamma^\mu (\partial_\mu + \frac{1}{2} \omega_\mu^{\; b c} J_{b c}) \psi \;.
\end{equation}
Now we are in the right starting position, to explore whether Weyl symmetry is indeed at work for real graphene.

\section{Fitting the theory in the lab}

A first trial is to check whether conformally flat $g^{(3)}_{\mu \nu}$ can make sense, i.e. what it is that we have to do to the two-dimensional graphene sheet so that the three-dimensional (time included) metric is conformally flat. This way, we can use the fact that the flat spacetime action is experimentally sound, and say that, through Weyl symmetry, it describes also the conformally flat physical situations. We are, of course, aware that for this to make sense for real graphene, the material should retain its Dirac-like and massless properties even in the deformed case. This is currently under investigation, by various researches, both at the theoretical and the experimental level. From the theoretical research, one learns that the two-sites-per-unit-cell structure of the honeycomb lattice of Fig.~\ref{honeycomb}, is resistent to deformation. Since this is the key topological property of the lattice responsible for the Dirac structure, it means that the latter is resistent to curvature, both intrinsic and extrinsic. From the experimental research, the fact that it is impossible to open-up a gap by only deforming the material (i.e. not including impurities), means that the masslessness is also a robust property, against deformations.

To have a conformally flat metric, since we are in three dimensions, the Cotton tensor must be zero (see, e.g., \cite{cs3})
\begin{equation}
    C_{\mu \nu} \equiv \epsilon_{\mu \lambda \kappa} \nabla^\lambda {R^{(3)}}^\kappa_\nu
    + \epsilon_{\nu \lambda \kappa} \nabla^\lambda {R^{(3)}}^\kappa_\mu = 0 \label{cotton} \;.
\end{equation}
What was found in \cite{iorio} is that all surfaces of constant Gaussian curvature $\cal K$ give rise to conformally flat (2+1)-dimensional spacetimes $g^{(3)}_{\mu \nu}$. Here a proof: The Ricci tensor is ${R^{(3)}}_\mu^\nu = {\rm diag}(0, {\cal K}, {\cal K})$. Nonzero components of Cotton: $C^{0 x} = - \partial_y {\cal K} = C^{x 0} $ and $C^{0 y} = \partial_x {\cal K} = C^{y 0}$, {\it q.e.d.}.

This result opens a window on a very rich scenario. What needs to be done to have conformally flat metrics, is just to have a surface of constant curvature. The (theoretical) simplicity of this operation, brings us one step closer to take advantage of this large symmetry that makes the flat space results (often exact) the ones to refer to, in order to have exact expressions for the full curved space results (an instance that is very rare in QFT in curved spacetimes). So, for instance, when we deal with two-point functions (as is customary when computing the electronic local density of states (LDOS)), we could use the formulae (\ref{directgreen2}), or (\ref{reversegreen1}), or else (\ref{reversegreen2}), and we would know exact expressions for these functions, that otherwise are tractable only via perturbative methods. But, as discussed around those equations, the issues to be solved, in order to be able to make physical sense of one or the other expression, are related to the process of measurement. In the case in point, this reveals to be a very difficult problem \cite{ioriolambiase,ioriolambiase2}, because we have to make sense of things like a comoving frame for graphene, or of what a $g^{(3)}_{00} \neq 1$ means for the measurements of time in a real laboratory. On top of that, we will have also to face very fundamental geometrical issues related to the embedding into $R^3$ (this refers to the purely spatial part of the story). Let me be a little more explicit.

The result (\ref{cotton}) is written in tensorial form, hence it is intrinsic. That is what is wanted in a true relativistic context. But here we are not, hence the frame of reference is, in general, very important for the experimental realization of configurations leading to observable effects. For instance, we need the explicit conformal factor to implement Weyl in the form outlined before, hence to know from (\ref{cotton}) that the metric is conformally flat is just the first step. We might need to change the frame
$q^\mu \equiv (t, x,y) \rightarrow {\cal Q}^\mu \equiv (T,X,Y)$ to see $g^{(3)}_{\mu \nu} ({\cal Q}) = \Phi^2 ({\cal Q}) \; g^{{\rm flat}}_{\mu \nu} ({\cal Q})$. First, this can be practically difficult. Second, and most important, we have to envisage the physical meaning of this coordinate frame, and we have to understand whether this is practically doable in a real experiment.

The ${\cal Q}^\mu$ for the sphere are not easy to envisage. On the other hand, all the infinite surfaces with ${\cal K} <0$ are better candidates, as the spatial part of the metric of graphene can be written, in isothermal coordinates, as
\begin{equation}
d\ell^2_{\rm graphene} = \frac{r^2}{{\tilde y}^2}(d{\tilde x}^2+d{\tilde y}^2) \label{lobsurface}\;,
\end{equation}
where ${\tilde x}, {\tilde y}$ are the \textit{abstract coordinates} of the Lobachevsky geometry in the upper half-plane (${\tilde y}>0$) model and $r=\sqrt{-{\cal K}^{-1}}$. By writing the full line element as
\begin{equation}
ds^2_{\rm graphene}=\frac{r^2}{{\tilde y}^2}\left[\frac{{\tilde y}^2}{r^2}dt^2-d{\tilde x}^2-d{\tilde y}^2\right] \label{lobspacetime}\;,
\end{equation}
the line element in square brackets is flat. This apparently solves our problem: the coordinates ${\cal Q}^\mu$ appear to be $(t, {\tilde x}, {\tilde y})$, as there we shall always have the explicit conformal factor $r^2/{\tilde y}^2$ to implement the Weyl symmetry. But it is not so, until we identify the specific surface and reduce to coordinates measurable in the Euclidean space $R^3$ of the laboratory! These matters have been explained in \cite{ioriolambiase}, and will be extensively presented in a forthcoming publication \cite{ioriolambiase2} (see also \cite{conference2011}). Let me give here the main arguments.

The actual type of surface is obtained by specifying ${\tilde x}$ and ${\tilde y}$ in terms of coordinates measurable using the Euclidean distance (embedding). This is crucial to describe the physical graphene spacetime. First of all, for all these surfaces, Hilbert's theorem holds: {\it It is impossible to have a complete surface of constant negative Gaussian curvature embedded in ${\bf R}^3$} (complete $\simeq$ no singularities). This will give raise to various singularities, that will have to be taken into account when analyzing the spacetime. Furthermore, there is the issue of having globally well defined conformal factor, hence expressions for the Green's function with global predicting power (all over the surface, excluding, perhaps the singular parts). This is not so for all surfaces of the family described by (\ref{lobsurface}). For instance, the elliptic pseudosphere \cite{spivak}
\begin{equation}
d{\ell}^2 = du^2 + r^2 \sinh^2 u/r dv^2 \label{elliptic} \;,
\end{equation}
with $u$ ($v$) meridian (parallel) coordinate, does not fit the scheme, in the sense that the conformal factor, $r^2/{\tilde y}^2$, is multivalued. Its natural coordinate system is not $q^\mu$, where the time is the lab time $t$, but, instead, a different one
\begin{equation}
{\cal Q}^\mu \equiv (T= e^{t/r} \cosh u/r, X= e^{t/r} \sinh u/r \cos v, Y= e^{t/r} \sinh u/r \sin v) \;.
\end{equation}
For those coordinates, though, it comes the problem of the practical realization of a frame of reference of this sort in a real experiment.

It should be clear from the above that the identification of the right surface is quite an art.

\section{Beltrami pseudosphere and Hawking effect}

As shown in \cite{ioriolambiase}, the Beltrami pseudosphere\footnote{Historically, this pseudosphere has been the first example of a surface of constant negative curvature. It convinced the mathematicians of the 1890s, that the exotic scenarios of the Lobachevsky geometry were possible to make real. It is said in many places, see, e.g., \cite{penrose}, that this surface does not deserve much admiration as, after all, it only conveys a little portion of the full intricacies of the Lobachevsky geometry. This latter instance is true, of course, and the Hilbert theorem above is just an effect of that. Nonetheless, I found this surface (and the many many others that came afterwards) a true piece of wander. To me it is like the intrusion into this (Euclidean, as for space) reality of a different reality, the border between them being the singular boundary (which I often call a ``Hilbert horizon''), beyond which the surface ceases to exist {\it here} ($z$ becomes imaginary), and continues {\it there}.}
\begin{equation}
d{\ell}^2 = du^2 + r^2 e^{2 u/r} dv^2
\end{equation}
with $v\in [0, 2\pi]$, $u\in [-\infty, 0]$, solves all those problems at once:

It gives raise to a three-dimensional conformally flat spacetime. In the frame of reference $q^\mu = (t,u,v)$, where the time is the lab time $t$, the conformal factor is such that the predictability is valid globally. Hence, through Weyl symmetry, we are able to produce (possibly exact) formulae, that hold all over the surface, and for any time.

The full line element is
\begin{equation}
ds^2_{\rm graphene} \equiv ds_{(B)}^2 = e^{2u/r} \left[e^{-2u/r}(dt^2-du^2)-r^2dv^2\right] \;,
\end{equation}
and we see that, we have an extra bonus: the line element in square brackets is the Rindler line element. This does not come as a surprise, if one looks at (\ref{lobspacetime}), where in square brackets we see {\it what would be a Rindler line element if the tilde coordinates were actually real, measurable coordinates}. It comes as a surprise, though, if one considers the argument given above about global validity of the coordinates and reference frames practically realizable in a laboratory (for the Beltrami, we are saying that all one needs to do is to curve the graphene sheet in that fashion). This does not mean that for the other surfaces of negative curvature we shall not be able to see any of the effects we are going to describe now for the Beltrami. It means, though, that, without changing coordinates, we shall be able to import the Beltrami results (valid there over the whole surface) only in a small neighbor of the surface, and that we shall not be able to infer from there what is going to happen in far parts of the surface. This might appear mind-twisting, but it is just a practical realization of what are the typical {\it Gedanken} experiments of QFT in curved spacetime, brought on the laboratory table.

Now, let us make use of this Rindler line element. We use the customary results of QFT in curved spacetimes, relating different observers to different quantum vacua (an instance introduced earlier here). We have the graphene frame, and we need an inertial frame to give meaning to the operational \textit{measuring procedure}. We model that by requiring the quantum vacuum of reference to be always the Minkowskian (inertial) one
$|0_M\rangle$. So, the pseudoparticles live in a curved spacetime, the measuring apparatus, by following the profile of the surface, as well (remember that the lab time $t$ is a Rindler time, see later). But the measuring apparatus cannot truly leave in a curved spacetime, there must be a place in the model where we stick in the information that the lab is in an inertial frame (no gravity!). This is done in \cite{ioriolambiase} by ascribing the inertial part of the measurement to the quantum vacuum of reference, hence chosen to be $|0_M\rangle$. The full explanation of this choice is in \cite{ioriolambiase}, and in the forthcoming \cite{ioriolambiase2}. Here let me just add to the above that, with this choice, we also try to take into account the fact that the pseudoparticles, once they leave the lattice, will ``collapse'' all their relevant Dirac properties, and will ``become'' again standard electrons.

\begin{figure}
\centering \leavevmode \epsfxsize=10cm \epsfysize=10cm
\epsffile{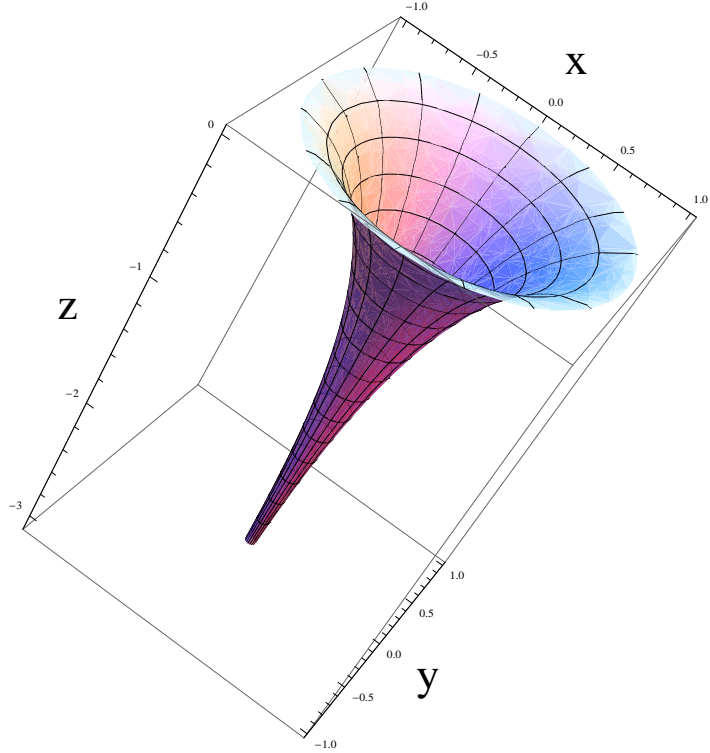}
\caption{The ${\bf R}^3$ coordinates of the Beltrami pseudosphere, in the canonical form, are \cite{spivak} $x(u,v) = R(u) \cos v$, $y(u,v) = R(u) \sin v$, $z(u) = r (\sqrt{1 - R^2(u)/r^2} - {\rm arctanh}\sqrt{1 - R^2(u)/r^2})$, with $R(u) =  c \, e^{u/r}$, $c > 0$ and $r = \sqrt{-{\cal K}^{-1}} > 0$ where ${\cal K}$ is the constant negative Gaussian curvature.  We can choose $c=r$, thus $R(u) \in [0,r]$ as $u \in [-\infty, 0]$. The surface is not defined for $R>r$ ($z$ becomes imaginary). In the plot $r=1$ and $u \in [-3.37, 0]$, $v\in [0, 2\pi]$.} \label{Beltrami}
\end{figure}

The positive frequency Wightman 2-point function to consider is then
\begin{equation}
    S^{(B)}(q_1, q_2)\equiv \langle 0_M|\psi^{(B)}(q_1){\bar \psi}^{(B)}(q_2)|0_M\rangle \,.
\end{equation}
Through Weyl, $g_{\mu\nu}^{(B)} = \varphi^2(u)  g_{\mu\nu}^{(R)}$, $\psi^{(B)} = \varphi^{-1}(u) \psi^{(R)}$, with $\varphi (u) = e^{u/r}$, hence
$S^{(B)}(q_1, q_2)=\varphi^{-1}(q_1) \varphi^{-1}(q_2)S^{(R)}(q_1, q_2)$, where $S^{(R)}\equiv \langle 0_M|\psi^{(R)}(q_1) {\bar \psi}^{(R)}(q_2)|0_M\rangle$ is the Rindler Green's function. The metric $g_{\mu\nu}^{(R)}$ is that of a flat {\it fictitious} spacetime (but Weyl-equivalent to the real Beltrami spacetime) in the {\it real} coordinates $q^\mu$. The physical result is recovered by simply multiplying the fictitious result by the proper factors.

The fictitious spacetime's line element is $ds_{(R)}^2 =  e^{-2u/r}(dt^2-du^2)-r^2dv^2$. We can use Minkowski coordinates (somehow hidden in the vacuum),  $T=r e^{-u/r}\sinh \frac{t}{r}$, $X=rv$, $Y=r e^{-u/r}\cosh \frac{t}{r}$, so that $Y^2-T^2=r^2 e^{-2u/r}\equiv \alpha^{-2}(u)$ for constant $u$ correspond to worldlines of observers moving at constant ``proper acceleration''
\begin{equation}
    \alpha(u)\equiv \frac{e^{u/r}}{r}\in [0, r^{-1}=\sqrt{-{\cal K}}] \;.
\end{equation}
This spacetime differs from standard Rindler spacetime in the following aspects i) a maximal acceleration, $\alpha_{\rm max}\equiv\alpha(u=0)=r^{-1}$, ii) confinement to one Rindler wedge ($\alpha \geq 0$), and iii) confinement on one side of the ``Hilbert horizon''  ($u \leq 0$) as the Beltrami world ends at $R=r$.

Introducing the customary Rindler coordinates, $\eta = t/r = \alpha_{\rm max} t$, $\xi = r e^{-u/r}=\alpha^{-1}(u)$, $ds^2_{(R)}=\xi^2 d\eta^2-d\xi^2-r^2dv^2$, we see that worldlines of constant proper acceleration in the fictitious Rindler spacetime correspond to a fixed point $({\bar u}, {\bar v})$ on the real Beltrami, with running time
\begin{equation}
    \frac{t}{r}\equiv \alpha({\bar u})\tau\,, \quad u={\bar u}\,, \quad v={\bar v} \;,
\end{equation}
so, to see the Unruh means to stay at a fixed point $({\bar u}, {\bar v})$ and ``wait'' till the horizon is reached, $t_{\rm hor} \sim r/v_F$. For the typical curvature we shall consider (we have to refer to the lattice spacing: $r>>\ell \simeq 2.5${\AA}) we have typical waiting times of the order of $t_{\rm hor} \sim 10^{-9}$s for $r \sim 1$mm, or $t_{\rm hor} \sim 10^{-12}$s for $r \sim 1\mu$m. The other time scale here is $\varepsilon$, i.e. the size in ``natural units'' of the detector. Thinking of an STM needle or tip, we have: for a tungsten needle $\varepsilon \sim 0.25{\rm mm} \times v_F^{-1} \sim 10^{-10}$s, while for a typical tip $\varepsilon \sim 10 {\rm {\AA}} \times v_F^{-1} \sim 10^{-15}$s.

Thus, the conditions are: consider $S^{(R)}(\tau, {\bf q}, {\bf q})$, where $\tau \equiv t / e^{{\bar u}/r}$, and stay at each point for the largest among $\varepsilon$ and the $t_{\rm hor}$. This is doable!

The {\it power spectrum}
\begin{equation}
F^{(R)}(\omega, {\bf q}) \equiv
\frac{1}{2} {\rm Tr} \left[\gamma^0\int_{-\infty}^{+\infty}d\tau e^{-i\omega \tau} S^{(R)}(\tau, {\bf q}, {\bf q})\right] \;,
\end{equation}
a part from inessential numerical factors, coincides here with the (not yet physical!) electronic LDOS  $\rho^{(R)}(\omega, {\bf q}) \equiv g/\pi F^{(R)}(\omega, {\bf q})$, where $g=4$ is the degeneracy. As we are in a massless case, this quantity can be computed exactly \cite{takagi}
\begin{equation}
    F^{(R)}(\omega)=\frac{1}{2}\, \frac{\omega}{e^{\omega/{\cal T}}-1} \;,
\end{equation}
where ${\cal T}\equiv \alpha({\bar u}) / (2\pi) =  e^{{\bar u}/r} / (2\pi r)$. Note the Bose-Einstein distribution, due to the (theoretically predicted) phenomenon known  as ``statistical swapping'' \cite{takagi}.

Going back to the physical spacetime is easy: $\rho^{(B)}(\omega, {\bf q}) = \varphi^{-2}({\bf q}) \rho^{(R)}(\omega, {\bf q})$. Hence, the LDOS we predict for a graphene sheet shaped as a Beltrami pseudosphere is (dimensional units are re-introduced)
\begin{equation}
 \rho^{(B)}(E, {\bar u}, r)= \frac{4}{\pi} \frac{1}{(\hbar v_F)^2} \frac{E \; e^{-2{\bar u}/r}}{\exp{\left[ E / (k_B {\cal T}_0 e^{{\bar u}/r}) \right]}-1} \label{ldosfinal} \;,
\end{equation}
where the temperature is a Hawking temperature
\begin{equation}\label{tzero}
    {\cal T}_0 \equiv \frac{\hbar v_F}{k_B 2 \pi r} \;.
\end{equation}
Note that the exact $\rho^{(B)}$ does not reduce to $\rho^{({\rm flat})} = \frac{2}{\pi} \frac{1}{(\hbar v_F)^2} |E|$ for $r \to \infty$ (${\cal K} \to 0$). This would be the case for a perturbative computation.

Currently, we are interacting with experimentalists towards a, direct or indirect, experimental detection of the behavior (\ref{ldosfinal}). The preliminary results of the theoretical setting-up for the experiment indicate that it should be possible to use more generic negatively curved surfaces. The logic is somehow the reverse, of what explained earlier about the impossibility to have globally valid predictions, for surfaces rather than the Beltrami, within the reference frame of the laboratory, and by acting solely on the geometry of the surface (i.e., not considering external electromagnetic fields, or other interactions). If the approach we described here is valid, due to the local isometry between every surface of constant negative curvature, a sort of Hawking-Unruh effect should be visible on any portion of those surfaces. It is matter of using the isometry (that is a change of {\it spatial} coordinates) in a small neighbor, and of knowing that the results cannot be extended beyond that neighbor. Nonetheless, an adapted form of the prediction (\ref{ldosfinal}) should be at work. These matters will be made more explicit in forthcoming publications.

\section{A ``quantum gravity'' desk-top lab?}

This research is the merging of various branches, but it is carried out mainly as an attempt to reconstruct in a laboratory a system as close as possible to what is believed to be a quantum field in a curved spacetime. The obvious ``classification'' of this type of activity would be under the class of ``analogue gravity''. In a way it is, of course, correct to say so. In another way, though, the analogies on the table here are many more than for other analogue systems. Here we have: i) a quantum pseudo-relativistic field description, ii) the possibility to see the effects of gravity emerging, in the form of curvature of the spacetime, iii)  a low-dimensional setting, that points to the use of exact results, both on the field theory side, and on the gravity side, not least the Weyl symmetry that points towards the importance of conformally flat spacetimes.

This last instance, immediately suggests scenarios like the conformally flat black hole space time (of constant negative curvature), discovered by Ba\~{n}ados, Teitelboim and Zanelli (BTZ) \cite{btz}. Some attempts towards trying to identify the physical conditions to reproduce such a situation on graphene were already mentioned in the first version of the arXiv entry \cite{ioriolambiase}. More recently, following the same general assumptions illustrated here, in \cite{Cvetic:2012vg} it was actually found how to relate the BTZ on graphene to a different pseudosphere, namely the hyperbolic pseudosphere with $d{\ell}^2 = du^2 + r^2 \cosh^2 u/r dv^2$. We are also currently investigating the consequences of this latter proposal \cite{weylBTZgraphene}.

In the title of this Section I use the words ``quantum gravity'', with an abuse of language (mitigated by the quotation marks), because I want to indicate that graphene has the potential to become a real laboratory to test fundamental ideas that go beyond the mere ``analogue gravity''. Indeed, as indicated in \cite{iorio}, many structures, typical of conformal field theories, arise in this contest. For instance, Virasoro and Liouville structures point to the possibility to test here (A)dS/CFT type of ideas. The amount of symmetries here, due to the dimensionality, and due to the lattice structure, is so high that the Hawking effect, although of great importance on its own, could truly be just the beginning of a series of investigations of fundamental ideas of nature realized in a laboratory. A natural next goal, after the experimental testing of the Hawking effect, could be the understanding of the ``black-hole thermodynamics'', starting from entropy and relative holography.

This far as for speculations, but there are various steps to be taken before getting anywhere near these goals. A crucial one is on the condensed matter side of the enterprise, and that is a solid geometrical/relativistic description of the elastic properties of the graphene membrane. It is matter there to have in the game the other actor of the story, i.e. the $\sigma$-bonds, and that is the other actor of the story also on the theory side, as this would tell us what sort of three-dimensional gravity theory is hiding here. There are also purely theoretical (if not entirely mathematical) questions to be fully addressed. The essential point, for this part, is that these spacetimes, e.g. what we called all along the Beltrami spacetime, are the result of fitting the models into the real laboratory, that, spatially, is $R^3$. Now, the subtleties are everything in delicate constructions like the BTZ black-hole. It is clear, for instance, that only certain global identifications will make a simple $AdS_3$ spacetime a true black hole \cite{BHTZ}. But even the standard $AdS_3$, to start with, is an issue here, because there is no such a thing as a spacetime with signature
$(+,-,-,+)$ where to embed the $AdS_3$ avoiding ``Hilbert horizons'' (i.e., the singular boundaries that we expect all the time from the Hilbert theorem). ``Real $AdS_3$'' deserve a thoroughly study, that gives them the status of proper spacetimes, not of accidents due to a wrong choice of coordinates. If not for other reason, this should be done even simply because this is what we can actually make in our laboratories. A last side remark is fully on the mathematics. It is an interesting problem to find the negative curvature  ``object'' corresponding to the icosahedron.

\section*{ACKNOWLEDGMENTS}

It is a pleasure to thank Ivo Sachs and Siddhartha Sen for the invitation, and Gaetano Lambiase for the enduring collaboration.

\end{document}